\documentstyle[epsfig]{l-aa} 
\font
\capt=cmti8.tfm    
\begin{document}
\title {Torque-luminosity correlation and possible 
evidence for core-crust relaxation in the X-ray pulsar GX 1+4}
\author{B. Paul, A.R. Rao, and K.P. Singh}
\institute {Tata Institute of Fundamental Research\\Homi Bhabha Road, Mumbai(Bombay) 400 005, India}
\offprints { B. Paul, {\it bpaul@tifrvax.tifr.res.in}}
\date{Received October ; accepted , 1996}
\thesaurus{13.25.5; 08.16.7 GX 1+4}
\maketitle
\markboth{ Paul et al.  : Torque-luminosity correlation in GX 1+4 }{}

\begin{abstract}
We present the detection of a positive correlation between spin-down 
rate $\dot{P}$ and pulsed X-ray luminosity in the BATSE archival data of 
the bright hard X-ray pulsar GX 1+4.  We have also seen a delay of 
5.6 $\pm$ 1.2 days between the luminosity change and the corresponding 
change in the spin-down rate. The observed correlation between
$\dot{P}$ and L$_X$ is used to reproduce the period history of GX 1+4 based on
the observed luminosity alone, and it is found that the spin period can be
predicted correct to 0.026\% when the luminosity is adequately sampled. 
The idea that at a higher luminosity more matter is accreted
and the accretion disk extends closer to the neutron star
thereby transferring more angular momentum to the system, seems not to be 
the case with GX 1+4.
The observed lag between the spin-down rate and the luminosity is
reported here for the first time in any such binary X-ray pulsar, and 
is found to be consistent with the time scale for the core-crust relaxation 
in a neutron star.


\end{abstract}

\section{\bf Introduction}

Period variations in X-ray binary pulsars are quite common and a number of 
pulsars show both spin-down and spin-up episodes over time scale of years 
or less. In binary systems with Roche-lobe overflow of the mass losing
secondary, such variations are generally explained in terms of the conventional 
accretion disk theory where the spinning-up or spinning-down of a neutron 
star of a given magnetic moment, mass and period depends only on the 
X-ray luminosity.
In binary systems containing massive early type secondaries, however,
accretion onto the neutron star is mostly through strong
stellar wind and conditions for forming stable accretion disks are
generally not present. According to the numerical simulations of 
mass accretion onto such systems (Taam \& Fryxell 1988; Blondin et al.
1990; Matsuda et al. 1991) the small accreted specific angular
momentum can change sign in an erratic manner which may lead to
alternating spin-up and spin-down episodes. Study of torque-luminosity
relationships in X-ray binary systems can therefore, be very 
instructive in understanding the accretion process in them.

The luminous hard X-ray pulsar GX 1+4, first detected in 1970 (Lewin et al.
1971), has several characteristics which makes it an ideal source to test
out the concepts of accretion powered X-ray pulsars.
 It has shown a continuous decrease of pulse period (spin-up) from about 135s 
in 1970 to about 110s in 1980 and it was included as one of the test sources 
in understanding the behaviors of disk-fed X-ray pulsars 
(Ghosh \& Lamb 1979a,b).
The source was below the detection limit of EXOSAT 
in 1983 (Hall \& Davelaar 1983) and 
after its rediscovery by GINGA in 1987 (Makishima et al. \cite{maki:a})
it has been showing a monotonically increasing spin period (spin-down),
except for a brief spin-up episode in between
(Finger et al. \cite{fing:a}; Chakrabarty et al. \cite{chak:a}).
GX 1+4 has been identified with a red giant M6III star V2116 Oph 
having an emission line spectrum that resembles a symbiotic star with a 
strong stellar wind (Davidsen et al. 1976). 
So far no binary period has 
been detected from this system, although optical pulsations 
with the same period as in X-rays
have recently been reported (Jablowski et al. 1996).
The presence of a giant
companion and period change sign reversals could imply that GX 1+4 is
a wind-fed system without a stable accretion disk. 
A correlation between spin-down and X-ray luminosity was pointed
out by Chakrabarty (1996), which is in apparent contradiction with the
general ideas of X-ray pulsars with accretion disks. 
To confirm the correlation between spin-down and X-ray luminosity found
by Chakrabarty (1996) and to understand the torque-luminosity relation
in greater detail, we have obtained the pulse period and luminosity history
of GX 1+4  for about 1200  days from the Compton Gamma Ray Observatory 
Science Support Center (COSSC) BATSE archive and carried out our analysis.
In the following sections, we present the analysis, results and its 
implications. 
\begin{figure*}[t]
\capt
\vskip 5.5cm
\includegraphics{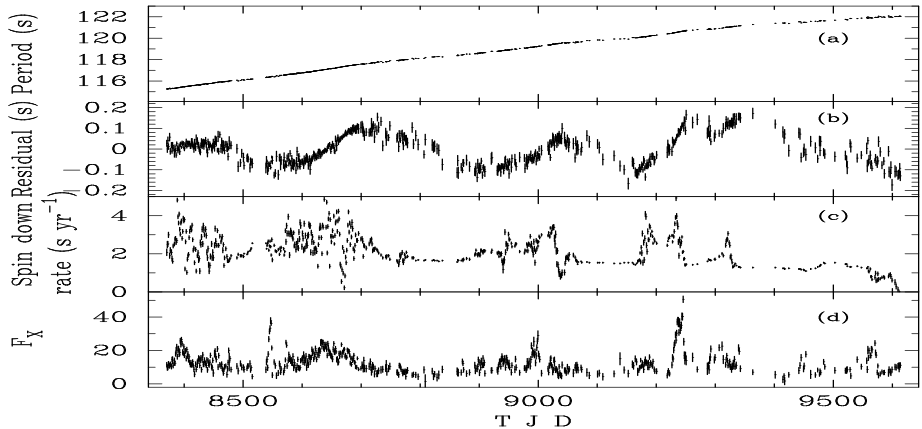}
Š\caption[]{ The pulse period  and the  pulsed flux history of GX 1+4 obtained
from the BATSE archive. The ordinate is time in Truncated Julian Days (TJD). 
a)  The  pulse  period b) a best fit quadratic function 
  subtracted from the period data to show  the  few 
hundred days features,
c) the instantaneous spin down rate $\dot{P}$ (see text) and d) The 
observed pulsed
flux F$_X$ at 40 keV (10$^{-5}$ photons cm$^{-2}$ s$^{-1}$ keV$^{-1}$). }
\end{figure*}

\section{\bf Data}
The pulse period and the luminosity of GX 1+4 for Truncated Julian 
Day (TJD) 8370 to 9615 (ie., 1991 April 24 to 1994 September 20)
were obtained from the COSSC archive. The 
observations were done with BATSE  Large Area Detectors (see Fishman 
et al. \cite{fish:a} for a description of BATSE). The archived data 
consists of the pulsar frequency obtained from blind searches and epoch 
folding performed on the BATSE data, the confidence level of the period 
determination and a Y/N flag indicating whether the confidence level exceeds 
a predetermined value. The archive also provides the pulsed flux F$_X$ at 
40 keV obtained by fitting an optically thin thermal bremsstrahlung (OTTB) 
spectrum of temperature 50 keV for channels 1 to 5 (25 to 98 keV). 
In the following analysis, we have taken only those data points with flag
Y (i.e., the period determination is reliable),  unless otherwise
mentioned.

\section{\bf Analysis and results}

The pulse period and the observed pulsed flux F$_X$ at 40 keV are plotted 
in Fig. 1 for TJD 8350 -- 9650 (Fig. 1a and 1d, respectively). A linear fit 
to the pulse period gives an average value for $\dot{P}$ of 2.12 s yr$^{-1}$. 
A quadratic fit to the data gives a value of $\dot{\nu}/\ddot{\nu}$ of 
8.3 yr. Higher order polynomials do not improve the fit. To see the 
period variation in more detail, the residuals to the quadratic fit are 
shown in Fig. 1b. The pulsed X-ray flux is seen to increase by 
a factor of $\ge$4 for duration of 2 $-$ 10 days compared to the average flux and 
by a factor of $\ge$2 for a duration of about 20 $-$ 100 days. 
Since F$_X$ is obtained after fitting a OTTB spectrum, it is in fact a 
measure of the hard X-ray pulsed luminosity. Though there have been some
indications of an anti-correlation between pulse fraction and total X-ray 
luminosity (Rao et al. \cite{rao:a}), the observed pulse fraction in the 
present spin-down era lies in a narrow range of 0.3 to  0.5. 
In fact, from a compilation of hard X-ray luminosity of GX 1+4 
(Chitnis \cite{chit:a}) we find a positive correlation between 
X-ray luminosity
and F$_X$. Hence, in the following, we treat F$_X$ as a 
measure of the total X-ray luminosity.
 
To examine whether the luminosity is related to the pulse period variation, 
the instantaneous spin down rate $\dot{P}$ is calculated for each of 
the data points  by doing a linear fit to the neighboring 25 data 
points and this is shown in Fig. 1c. The similarity in the Figs. 1c and 1d 
led to further analysis of correlation and cross-correlation between 
$\dot{P}$ and the pulsed flux F$_X$.

\subsection { Spin down rate and luminosity}

To estimate the correlation of spin change rate and 
luminosity we choose only those $\dot{P}$ values where the linear fit
around that data point (for about $\pm$ 12 days) is acceptable (unlike Fig. 1c.
where all the data points are included).
 As the regions of very high F$_X$ are of short durations, the determination
of $\dot{P}$ is not very reliable and hence
we exclude those points from our analysis. 
The two
quantities are positively correlated and we have calculated a 
correlation coefficient of 0.63 (for 102 data points) and the probability 
of no correlation in the given data set is estimated to be 10$^{-12}$.

To investigate 
whether the pulse period variation is completely governed by the
luminosity variation,
we made an attempt to reproduce the pulse period history of GX 1+4 only
from the luminosity history.
The positive correlation seen between $\dot{P}$ and F$_X$ 
is assumed to be the real torque
transfer equation in the pulsar and the pulse period of the first data
point is propagated with time depending on F$_X$, using the linear relation
obtained between $\dot{P}$ and F$_X$. For this purpose we have
used those F$_X$ values even when the period determination is uncertain
(Flag N in the archive). The resultant residuals in the period determination
are shown in Fig. 2b. For comparison, we show in Fig. 2(a) the residuals 
to the period obtained by assuming a constant $\dot{P}$.  
The 100 $-$ 200 days features in the upper plot 
is not present in the lower plot signifying
that the pulse period changes are actually correlated to the luminosity.
However, the reproduction of the pulse period for days later than TJD 9000
deviates from the observed one by up to about 0.5 s because of the lack of 
sufficient number of F$_X$ measurements. The rms deviation in the pulse
period as estimated from only a constant $\dot{P}$ (Fig. 2a)
is 0.1 s and it improves to 0.04 s when pulse period is predicted 
from the $\dot{P}$ $-$ L$_X$ relation (Fig. 2b). The rms deviation reduced
further to 0.03 s (which is the typical error in the period determination) 
for TJD 8370 to 9000 (where F$_X$ is well determined and well
sampled). Hence, we can conclude that when F$_X$ is well sampled, all the 
variations in period can be explained correctly within the observational 
errors using a simple linear relation between $\dot{P}$ and F$_X$.

\begin{figure}[t]
\vskip 5.5cm
\includegraphics{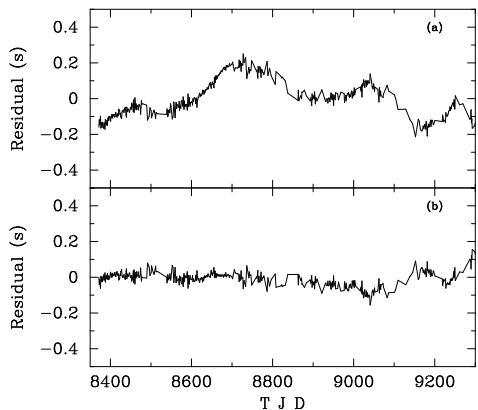}
\caption[]{ Residuals of pulse period plotted versus time in TJD.
a) The 100-200 days features in the
observed period when a linear fit is subtracted.  b)
The coefficients for a best fitted straight line between $\dot{P}$
and  F$_X$ is used to reproduce the pulse period of GX  1+4,
only taking pulsed luminosity into consideration and the difference
 of the observed
and calculated period is shown here. }
\end{figure}

\subsection { Time delay }

The instantaneous spin-down rate and the pulsed flux were subjected to
cross-correlation tests. For this purpose $\dot{P}$ is calculated using
two neighboring data points and the average value of F$_X$ is used.
When the total data is taken we find a positive correlation
between $\dot{P}$ and F$_X$  at a confidence level of 99.4\%.
The reduced level of confidence is due to the fact that $\dot{P}$
is calculated over 2 observations (unlike $\pm$12 data points used
in the previous section). The correlation, however, was found to be
delayed by a few days. To improve the confidence level,
the total data are divided into several sets of 128 data 
points and the derived cross-correlation values are co-added. The
resultant profile is shown in Fig. 3. The central part
of the figure is shown in an expanded form in the inset
to the figure. As can be seen from the figure, there is a clear 
asymmetry near 0. A Gaussian fit to the profile 
near 0 gives a $\chi^2$ of 20 for 35 degrees of freedom (dof) and the
derived value of  delay is 
(4.8 $\pm$ 1.0) $\times$ 10$^5$ s (5.6$\pm$1.2 days).
 The errors are calculated by the
criterion of $\chi_{min}^2$+2.3 (1 $\sigma$ error for two free
parameters). 
A constant fit to the profile gives $\chi^2 =$  
75 for 36 dof 
showing the existence of correlation at a confidence level of 99.99\%.
This confidence level improves further if the value of the
constant is kept fixed at 0 (i.e., 
there is no correlation instead of constant correlation).
A Gaussian fit with the centroid frozen at zero gives $\chi^2 =$
54  signifying the existence of a delay at a very high 
confidence level (the value of $\Delta \chi^2$ being 34 for one
additional parameter).
Hence the co-adding method resulted in the detection of a delay at a
high confidence level, and  could be 
the reason for the lack of detection of any
such delay by other workers (Chakrabarty 1996).
The delay between $\dot{P}$ and F$_X$ is seen for the first time in
an X-ray pulsar.
\begin{figure}[h]
\vskip 4.5cm
\includegraphics{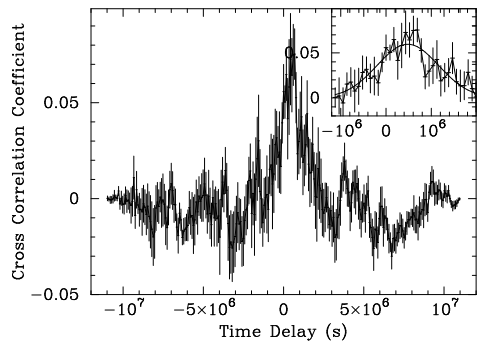}
\caption[]{ The cross correlation between $\dot{P}$ and F$_X$ is plotted here.
Inset  gives expanded view for the central part of the plot.  The asymmetry
seen here corresponds to 5.6$\pm$1.2 day delay for $\dot{P}$ compared to F$_X$.
}
\end{figure}

\section{\bf Discussion }

The frequent and large variations in the X-ray luminosity of GX 1+4
observed with BATSE are quite similar to those found in the other 
accretion-powered X-ray pulsars.  Power spectrum analysis of the
luminosity fluctuations shows a power-law component (index= $-$2.1) indicative
of a red noise in the system and has been seen before (Baykal \& Ogelman 1993).
Power spectrum analysis of the period fluctuations 
also shows a red-noise component.  Period fluctuations have
also been seen in other accreting pulsars with time scales down to a few days, 
but the red noise component seen here suggests that these fluctuations might
represent torques that are internal to the neutron star rather than due
to inhomogeneities in the accretion flow (White et al. 1995).
The luminosity fluctuations are found to be correlated with 
the instantaneous $\dot{P}$, which always stays positive throughout the observation.
The direct positive correlation of $\dot{P}$ 
with the X-ray luminosity is difficult 
to explain in terms
of accretion disk models (Ghosh \& Lamb 1991) as
in such models an increase in luminosity is related to increased mass
accretion rate that decreases the inner radius of the disk and leads to a 
spin-up of the neutron star i.e., negative $\dot{P}$. On the other hand,
if GX 1+4 accretes matter directly through stellar wind with negligible
specific angular momentum, then the reversal of spin change sign could mean
a reversal in the direction of the small disk that can form.
A positive correlation
between $\dot{P}$ and L$_X$ can then be expected, as a sudden decrease
in the net angular momentum can lead to an increase in accretion (King 1995).

The delay between L$_X$ and $\dot{P}$ is difficult to explain in any accretion
theory. The region of hard X-ray emission is very close to the neutron star
surface and one cannot expect any delay between the X-ray emission and the 
resultant angular momentum transfer to the neutron star. Hence we look
for some phenomena internal to the neutron star as a possible explanation
to the delay. In this regard it is very instructive to compare these results
to a similar phenomena observed in GRO J1744-28 (Stark et al. 1996).
Stark et al. have found a phase lag in the bursting X-ray pulsar GRO J1744-28. 
During an X-ray burst when the X-ray luminosity increased by more than a factor
of 15 in about 10 s, the phase lag increased to about 28 ms and subsequently
the phase lag relaxes back with an exponential decay time of about 720 s. 
Interpreting this phenomenon in terms of models for pulsar glitches developed
for radio pulsars, the phase lag during
the burst corresponds to an initial spin-down with $\Delta\Omega/\Omega \sim$
 $-$ 10$^{-3}$. The exponential decay time scale is equated to the crust-core
coupling time scale, which is (4$\times$10$^2$ $-$ 10$^4$) P, where P is the 
rotation period of the neutron star (Alpar \& Sauls 1988).
If the phenomena observed in GRO J1744-28 is treated as an impulse response
to luminosity change and if this phenomena is common to GX 1+4 too,
continuous changes in luminosity (as seen in GX 1+4) 
will reflect as a delay in the $\dot{P}$ variation.
The magnitude of period variation in GX 1+4 
(dP/P = $-$ $\Delta\Omega/\Omega$ $\sim$  10$^{-3}$) is comparable
to that seen in GRO J1744-28. 
Further, the observed time scale (6 days) agrees with the relation between
$\tau$ and   P given by Alpar \& Sauls (1988).

The observed lag of $\dot{P}$ with respect to L$_X$ is, therefore, 
consistent with the impulse response
of X-ray luminosity variation seen in GRO J1744-28, with the time scales
scaled up according to the relation given for core-crust relaxation.
As pointed out by Stark et al., for the core-crust relaxation to occur, 
first the crust has to decouple and the angular momentum has to be
transferred to the crust and the crust couples back to the core in a 
time scale given by Alpar \& Sauls. 

{\it In conclusion}, our analysis of the period and X-ray luminosity history
of GX 1+4 observed with the BATSE shows: 
(i) a positive correlation between pulsed hard X-ray luminosity and 
spin-down rate, and (ii) the spin-down rate lags by 5.6$\pm$1.2 days
with respect to the pulsed luminosity.  These results suggest that the internal
torque of the neutron star can play a dominant role in the period-luminosity
history of GX 1+4.   

\vskip 0.15 in

\begin{acknowledgements}
We thank the BATSE team and COSSC for providing the valuable pulsar data.
We thank the anonymous referees for their comments and suggestions.
\end{acknowledgements}

{}


\begin{thebibliography}{}

\bibitem[1988]{alpar:a}
Alpar, M. A., Sauls, J. A., 1988, ApJ 327, 723

\bibitem[1993]{baykal:a}
Baykal, A., Ogelman, H., 1993, A\&A 267, 119

\bibitem[1993]{blondi:a}
Blondin, J.M., Kallman, T.R., Fryxell, B.A., Taam, R.E., 1990, ApJ 356, 591.

\bibitem[1994]{chak:a}
Chakrabarty D., Prince T. A., Finger M. H., 1994, IAU Circular Nr. 6105

\bibitem[1996]{chak:c}
Chakrabarty D., 1996, Presented at the Symposium on {\it X-ray Timing},
31st COSPAR Scientific Assembly, 14 $-$ 21 July, 1996, Birmingham, UK.

\bibitem[1994]{chit:a}
Chitnis V. R., 1994, PhD Thesis, Bombay University.

\bibitem[1976]{davids:a}
Davidsen, A., Malina, R.F., Bowyer, S., 1976, ApJ 203, 448.

\bibitem[1993]{fing:a}
Finger M. H., Wilson R. Rb., Fishman G. J., et al., 1993 IAU Circular Nr. 5859

\bibitem[1989]{fish:a}
Fishman, G. J. et al., 1989 in Proc. of the GRO Science Workshop, 
ed. W. N. Jonshon (Greenbelt: NASA/GSFC), p2.

\bibitem[1979a]{pghosh:a}
Ghosh P., Lamb F. K., 1979a, ApJ 223, L83

\bibitem[1979b]{pghosh:b}
Ghosh P., Lamb F. K., 1979b, ApJ 234, 296

\bibitem[1991]{pghosh:c}
Ghosh P., Lamb F. K., 1991, in Neutron Stars: Theory and Observations, 
eds. J. Venturs, D. Pines, (NATO ASI Ser. C, 344) (Dordrecht: Kluwer), p363

\bibitem[1983]{hall:a}
Hall, R., Davelaar J., 1983, IAU Circular Nr. 3872

\bibitem[1996]{jablow:a}
Jablowski, F., Pereira, M., Braga, J., Campos, S.J., Gneiding, C.,
1996, IAU Circular Nr. 6489

\bibitem[1995]{king:a}
King, A., 1995, in X-ray Binaries,
eds. W. H. G. Lewin, Jan van Paradijs, E.P.J. van den Heuvel,
Cambridge University Press, Cambridge, p419

\bibitem[1971]{lewin:a}
Lewin, W. H. G., Ricker, G., McClintock, J.E., 1971, ApJL 169, L17

\bibitem[1988]{maki:a}
Makishima  K., Ohashi T., Sakao T., et al., 1988, Nat 333, 746

\bibitem[1991]{matsud:a}
Matsuda, T., Sekino, N., Sawada, K., et al., 1991, A\&A 248, 301.

\bibitem[1994]{rao:a}
Rao A. R., Paul B., Chitnis V. R., Agrawal P. C.,
Manchanda R. K., 1994, A\&A 289, L43

\bibitem[1996]{stark:a}
Stark, M.J., Baykal, A., Strohmayer, T., Swank, J.H. 1996, ApJ 470, L109 

\bibitem[1988]{taam:a}
Taam, R.E., Fryxell, B.A., 1988, ApJ 327, L73.

\bibitem[1995]{white:a}
White, N.E., Nagase, F., Parmar, A. N., 1995, in X-ray Binaries,
eds. W. H. G. Lewin, Jan van Paradijs, and E.P.J. van den Heuvel,
Cambridge University Press, Cambridge, p1.
\end{thebibliography}
\end{document}